\newcommand{\ket}[1]{| \, #1 \, \rangle}

\documentclass{svjour3}                     % onecolumn (standard format)
\smartqed  % flush right qed marks, e.g. at end of proof
\usepackage{graphicx}
\begin{document}

\title{Narrow bound states of the $DNN$ system%\thanks{Grants or other notes
}


\author{E. Oset         \and
        M. Bayar \and
	C. W. Xiao \and
	T. Hyodo \and
	A. Dote \and
	M. Oka
}


\institute{E. Oset \at
              Departamento de Fisica Teorica and IFIC, Centro Mixto Universidad de Valencia-CSIC,
Institutos de Investigacion de Paterna, Aptdo. 22085, 46071 Valencia, Spain
              Tel.: +34-96-3543525\\
              Fax: +34-96-3543488\\
              \email{oset@ific.uv.es}           %  \\
           \and
           M. Bayar \at
             Departamento de Fisica Teorica and IFIC, Centro Mixto Universidad de Valencia-CSIC,
Institutos de Investigacion de Paterna, Aptdo. 22085, 46071 Valencia, Spain;
 Department of Physics, Kocaeli University, 41380 Izmit, Turkey
\and 
C.W. Xiao \at 
Departamento de Fisica Teorica and IFIC, Centro Mixto Universidad de Valencia-CSIC,
Institutos de Investigacion de Paterna, Aptdo. 22085, 46071 Valencia, Spain 
\and
T. Hyodo \at
Department of Physics, Tokyo Institute of Technology, Meguro 152-8551, Japan
\and
A. Dote \at
High Energy Accelerator Research Organization (IPNS/KEK), 1-1 Ooho, Tsukuba, Ibaraki, Japan, 305-0801; J-PARC Branch, KEK Theory Center, Institute of Particle and Nuclear Studies,
High Energy Accelerator Research Organization (KEK),
203-1, Shirakata, Tokai, Ibaraki, 319-1106, Japan
\and 
M. Oka \at
Department of Physics, Tokyo Institute of Technology, Meguro 152-8551, Japan
}

\date{Received: date / Accepted: date}

\maketitle

\begin{abstract}
We report on a recent calculation of the properties of the $DNN$ system, a charmed meson with two nucleons. The system is analogous to the $\bar K NN$ system substituting a strange quark by a charm quark. Two different methods are used to evaluate the binding and width, the Fixed Center approximation to the Faddeev equations and a variational calculation.  In both methods we find that the system is bound by about 200 MeV and the width is smaller than 40 MeV, a situation opposite to the one of the $\bar K NN$ system and which makes this state well suited for experimental observation. 

\end{abstract}

\section{Introduction}
\label{intro}
The $\bar K NN$ system has been the subject of much attention and recent papers converge to having bindings of about 20 MeV and large widths of about 80 MeV \cite{Dote:2008hw,Ikeda:2010tk,Barnea:2012qa,Bayar:2011qj,Oset:2012gi,Bayar:2012rk}. A fraction of about 30 MeV of the width of the state of comes from absorption of the $\bar K$ on the pair of nucleons, recently evaluated with precision in \cite{melaabsorption}. The large width can be intuitively understood since the $\bar K N $ merges into a $\Lambda(1405)$ that has a width of about 30 MeV, but since the $\Lambda(1405)$ can be formed with either nucleon, the width can be estimated of the order of 60 MeV, to which the absorption \cite{melaabsorption} must be added. It is no wonder that with a width much larger than the binding such state has not been found in spite of searches and claims (see discussion in \cite{Oset:2007vu,Ramos:2007zz}). 

The fate of the analogous $DNN$ system could be quite different. Indeed, the analogous resonance, according to studies of the $DN$ interaction with coupled channels \cite{Hofmann:2005sw,Mizutani:2006vq}, is the $\Lambda_c(2595)$, which has a width of 2.6 MeV \cite{pdg}. On the other hand, the binding of the $DN$ system, that by analogy to the $\bar K N$ goes as the relativistic energy of the $D$, should also be bigger than in the case of the $\bar K N$ system. As a consequence we are led to have a state more bound and with a smaller width, which could be easily observable. In the study done in \cite{dnn} this is what is observed looking into the problem from two perspectives: one is using the Fixed Center approximation to the Faddeev equations and the other one using a variational calculation. Both methods converge into a common answer providing a state around 3500 MeV with a width of about 30-40 MeV counting the absorption of the $D$ by two nucleons.

\section{The Fixed Center Approximation}
\begin{figure}[tbp]
\centering
\includegraphics[width=0.90\textwidth]{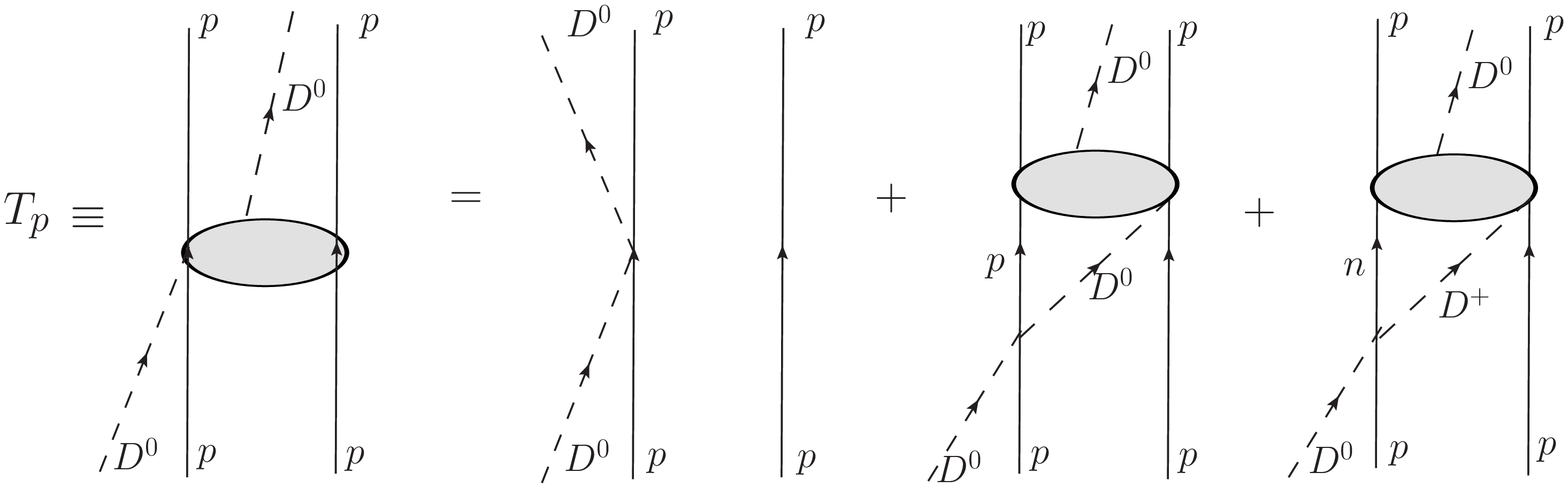}\\
\includegraphics[width=0.55\textwidth]{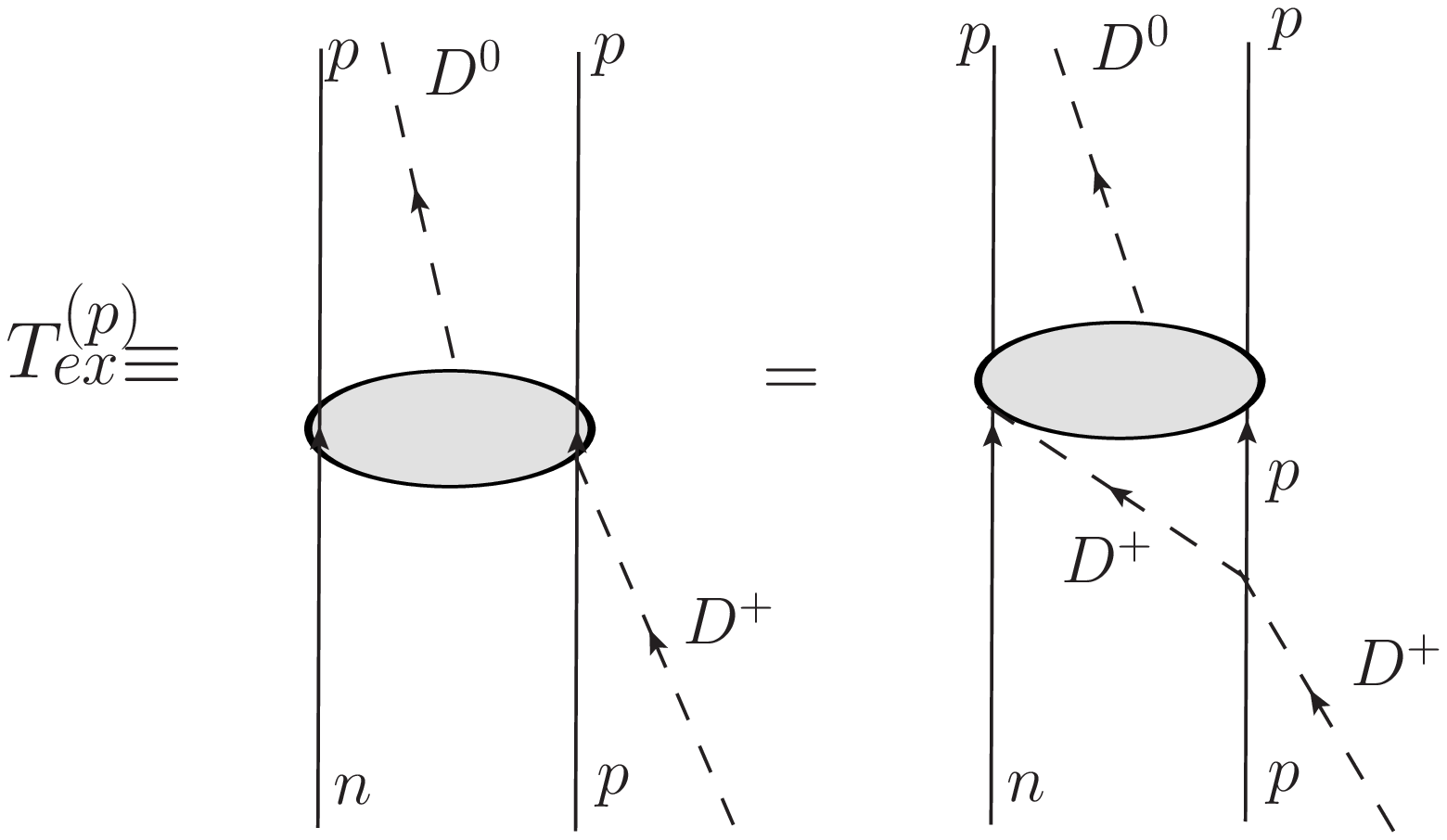}\\
\includegraphics[width=0.95\textwidth]{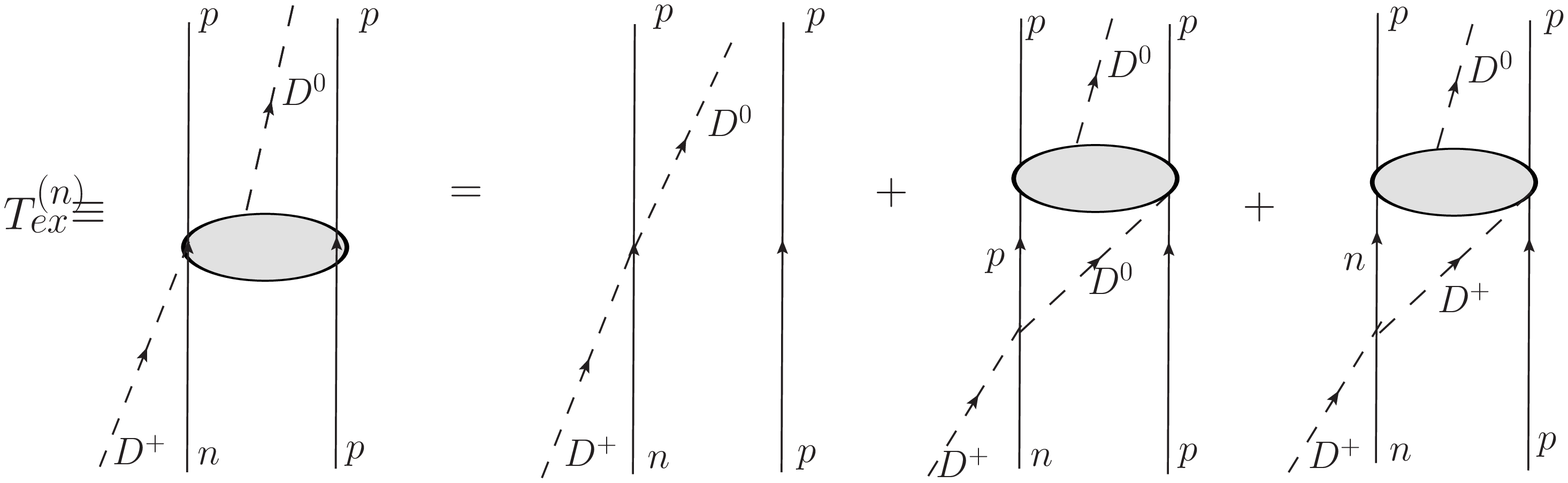}
\caption{Diagrammatic representations of the partition functions for the $D^0 p p \rightarrow D^0 p p $.}
\label{fig:tpp22}
\end{figure}

 In this approach we consider that the two nucleons form a cluster and that the $D$ scatters with these nucleons without changing them from their ground state. This is fair when one has a bound D, which has no energy to excite the $NN$ system. Under this assumption one has a set of coupled equations involving partition functions which for the $D^0 p p $ system sum the diagrams that we can see in Fig. \ref{fig:tpp22}.
 
\begin{figure}%[tbp]
\centering
\includegraphics[width=0.65\textwidth]{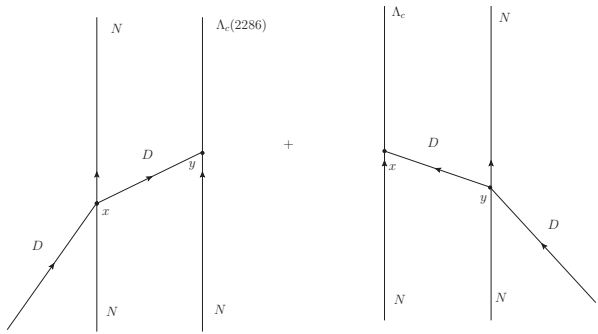}
\caption{Diagrammatic representation of the $D (N N)$ absorption.}
\label{FCAfignew}
\end{figure} 

These amplitudes fulfill a set of coupled equations 
\begin{eqnarray}
T_{p}&=&t_{p}+t_{p}G_0T_{p}+t_{ex}G_0T_{ex}^{(p)}\nonumber\\
T_{ex}^{(p)}&=&t_{0}^{(p)}G_0T_{ex}^{(n)}\nonumber\\
T_{ex}^{(n)}&=&t_{ex}+t_{ex}G_0T_{p}+t_{0}^{(n)}G_0T_{ex}^{(p)},
\label{Eq:tptexp}
\end{eqnarray}
where the two-body amplitudes are given as
$t_p=t_{D^0 p , D^0 p}$, $t_{ex}=t_{D^0 p , D^+ n}$, $t_{0}^{(p)}=t_{D^+ p , D^+ p}$, and $t_{0}^{(n)}=t_{D^+ n , D^+ n}$.  A set of similar, but easier equations, are obtained for scattering of the $DNN$ in isospin $I=3/2$ such that the proper $I=1/2$ amplitude where the bound $DNN$ state appears is given by

\begin{eqnarray}
T^{(1/2)}&=&\frac{\frac{3}{2}t^{(0)}+\frac{1}{2}t^{(1)}+2G_0t^{(0)}t^{(1)}}{1+\frac{1}{2}(t^{(1)}-t^{(0)})G_0-G_0^2t^{(0)}t^{(1)}},
\nonumber
\end{eqnarray}
where $t^{(0)}, ~t^{(1)}$ are the isospin $I=0,1$ $DN$ amplitudes and $G_0$ is the $D$ propagator folded by the $NN$ form factor.

\begin{equation}
G_0=\int\frac{d^3q}{(2\pi)^3}F_{NN}(q)\frac{1}{{q^0}^2-\vec{q}\,^2-m_{D}^2+i\epsilon} ,
\label{Eq:gzero}
\end{equation}

The absorption of the $D$ by two nucleons is based on the diagrams of Fig. \ref{FCAfignew}, and they are included in a nonperturbative way where the elementary $DN$ amplitudes are already modified to account by the possibility of the $D$ being absorbed by a second nucleon.

\begin{figure}
\centering
\includegraphics[width=0.8\textwidth]{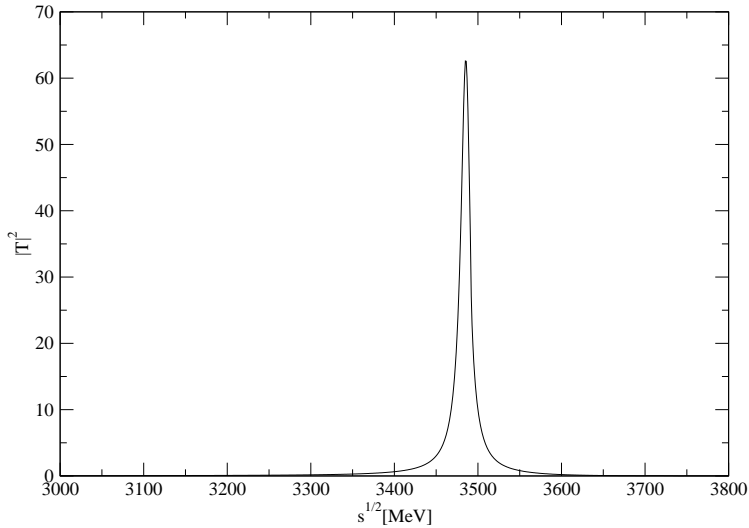}
\caption{Modulus squared of the three-body scattering amplitude for $I=1/2$ and $J=0$ including absorption with reduced $NN$ radius.}
\label{fig:redtmats0}
\end{figure}

The modulus squared amplitude for the $DNN$ amplitude is shown in Fig. \ref{fig:redtmats0}. We can see a clear peak around 3500 MeV with a width of about 20-30 MeV, which indicates the appearance of a state of the $DNN$ system.

\section{Variational calculation}
\label{sec:2}

Here we calculate the energy of the $DNN$ system with a variational approach formulated for the $\bar{K}NN$ system in Refs.~\cite{Dote:2008hw,Dote:2008in}. As in the case of the FCA, we consider the $DNN$ system with total isospin $I=1/2$ and the total spin-parity  $J^{P}=0^{-}$. The trial wave function for the state is prepared with two components:
\begin{equation}
    \ket{\Psi^{J=0}} 
    = (\mathcal{N}^{0})^{-1}[\ket{\Phi_{+}^{0}}
    +C^{0}\ket{\Phi_{-}^{0}}] ,
    \nonumber
\end{equation}
where $\mathcal{N}^{0}$ is a normalization constant and $C^{0}$ is a mixing coefficient. In the main component $\ket{\Phi_{+}^{0}}$, two nucleons are combined into spin $S_{NN}=0$ and isospin $I_{NN}=1$ so all the two-body subsystems can be in $s$ wave. We also allow a mixture of the $\ket{\Phi_{-}^{0}}$ component where both spin and isospin are set to be zero, so the orbital angular momentum between two nucleons is odd.

We consider the following Hamiltonian in this study:
\begin{equation}
    \hat{H}
    = \hat{T}+\hat{V}_{NN}+Re\hat{V}_{DN}
    -\hat{T}_{c.m.} ,
    \label{eq:Hamiltonian}
\end{equation}
where $\hat{T}$ is the total kinetic energy, $\hat{V}_{DN}$ is the $DN$ potential term which is the sum of the contributions from two nucleons, and $\hat{T}_{c.m.}$ is the energy of the center-of-mass motion. For the $NN$ potential $\hat{V}_{NN}$, we use three models: HN1R which is constructed from Hasegawa-Nagata No.1 potential~\cite{PTP45.1786}, the Minnesota force~\cite{Thompson:1977zz}, and the gaussian-fitted  version of the Argonne v18 potential~\cite{Wiringa:1994wb}.

The $DN$ potential in this approach is obtained by studying the $DN$ scattering in the coupled channels of \cite{Mizutani:2006vq} and eliminating all except for the $DN$ one with an effective potential such as to obtain the same scattering amplitude as with the coupled channels \cite{Hyodo:2007jq}. As in \cite{Mizutani:2006vq}, we consider seven (eight) coupled channels in the isospin $I=0$ ($I=1$) sector, $DN$, $\pi\Sigma_{c}$, $\eta\Lambda_{c}$, $K\Xi_{c}$, $K\Xi_{c}^{\prime}$, $D_{s}\Lambda$, and $\eta^{\prime}\Lambda_{c}$ ($DN$, $\pi\Lambda_{c}$, $\pi\Sigma_{c}$, $\eta\Sigma_{c}$, $K\Xi_{c}$, $K\Xi_{c}^{\prime}$, $D_{s}\Sigma$, and $\eta^{\prime}\Sigma_{c}$).

\begin{table}[tdp]
\caption{Results of the energy compositions in the variational calculation for the ground state of the $DNN$ system with total isospin $I=1/2$ (range parameter $a_{s}=0.4$ fm). Terms ``bound'' and ``unbound'' are defined with respect to the $\Lambda_{c}^{*}N$ threshold. All the numbers are given in MeV.}
\begin{center}
%\begin{ruledtabular}
\begin{tabular}{lrrrr}
                      & HN1R     &          & Minnesota & Av18 \\
                      & $J=1$    & $J=0$    & $J=0$     & $J=0$  \\
\hline
                      & unbound  & bound    & bound     & bound \\
$B$                   & 208      & 225      & 251       & 209 \\
$M_{B}$               & 3537     & 3520     & 3494      & 3536 \\
$\Gamma_{\pi Y_{c}N}$ & -        & 26       & 38        & 22 \\[5pt]
$E_{kin}$      & 338      & 352      & 438       & 335 \\
$V(NN)$               & 0        & $-2$     & 19        & $-5$\\
$V(DN)$               & $-546$   & $-575$   & $-708$    & $-540$ \\
$T_{nuc}$      & 113      & 126      & 162       & 117 \\
$E_{NN}$              & 113      & 124      & 181       & 113 \\[5pt]
$P(Odd)$       & 75.0 \%  & 14.4 \%  & 7.4 \%    & 18.9 \% \\
\end{tabular}
%\end{ruledtabular}
\end{center}
\label{tab:energy}
\end{table}%

In Table~\ref{tab:energy} we show some of the properties of the state found for different $NN$ potentials. As seen in the Table, the $DNN$ system in the $J=0$ channel is bound below the $\Lambda_{c}^{*}N$ threshold ($B\sim 209$ MeV) for all the $NN$ potentials employed. A large kinetic energy of the deeply bound system is overcome by the strong attraction of the $DN$ potential, while the $NN$ potential adds a small correction. Comparing the results with three different nuclear forces, we find that the binding energy is smaller when the $NN$ potential has a harder repulsive core.

Although we will not discuss it here, we also find in \cite{dnn} a state with $J=1$ but less bound and more uncertain than the $J=0$ that we have exposed.

\section{Possible experiments to produce the $DNN$ state}

As a suggestion to observe experimentally this state we can think of the $\bar{p}{~^3}He \rightarrow \bar{D}^0 D^0pn\to \bar{D}^{0} [DNN]$ reaction, which could be done by FAIR at GSI. With a $\bar{p}$ beam of $15~GeV/c$ there is plenty of energy available for this reaction and the momentum mismatch of the $D^0$ with the spectator nucleons of the $^3$He can be relatively small. Estimations made in \cite{dnn} indicate that one would expect several thousand events per day for the background of the proposed reaction. A narrow peak could be visible on top of this background corresponding to the $DNN$ bound state formation.

Another possibility is the high-energy $\pi$ induced reaction. An analogous reaction is $\pi^{-} d\to D^{-}D^{+}np \to D^{-} [DNN]$ where the relevant elementary process is $\pi^{-}N\to D^{+}D^{-}N$. Since the $DN$ pair in the $DNN$ system is strongly clustering as the $\Lambda_{c}^{*}$, the reaction $\pi^{-} d\to D^{-}\Lambda_{c}^{*}n \to D^{-} [DNN]$ is also another candidate. The elementary reaction $\pi^{-}p\to D^{-}\Lambda_{c}^{*}$ is  relevant in this case. Such reactions may be realized in the high-momentum beamline project at J-PARC.

 \section{Conclusions}
%%%%%%%%%%%%%%%%%%%%%%%%%%%%%%%%%%%%%%%%%%%%%%%%%%%%%%%%%%%%%%%%%%%
  
We have studied the $DNN$ system with $I=1/2$ using two independent methods: the Fixed Center Approximation to the Faddeev equations and a variational method, and have found that the system is bound and rather stable, with a width of about 20-40 MeV. We obtained a clear signal of the quasi-bound state for the total spin $J=0$ channel around 3500 MeV.

The small width of the $DNN$ quasi-bound state is advantageous for the experimental identification. The search for the $DNN$ quasi-bound state can be done by $\bar{p}$ induced reaction at FAIR, $\pi^{-}$ induced reaction at J-PARC, and relativistic heavy ion collisions at RHIC and LHC.

\section{Acknowledgments}
This work is partly supported by  projects FIS2006-03438 from the Ministerio de Ciencia e Innovaci\'on (Spain), FEDER funds and by the Generalitat Valenciana in the program Prometeo/2009/090.
This research is part of the European
 Community-Research Infrastructure Integrating Activity ``Study of
 Strongly Interacting Matter'' (acronym HadronPhysics2, Grant
 Agreement n. 227431) 
 under the Seventh Framework Programme of EU. 
T.H. thanks the support from the Global Center of Excellence Program by MEXT, Japan, through the Nanoscience and Quantum Physics Project of the Tokyo Institute of Technology. 
This work is partly supported by the Grant-in-Aid for Scientific Research from 
MEXT and JSPS (Nos.\
  24105702
  and 24740152).

%\begin{acknowledgements}
%\end{acknowledgements}



\end{document}